\documentclass{article}

    \usepackage[preprint]{neurips_2025}

\usepackage[utf8]{inputenc} 
\usepackage[T1]{fontenc}    
\usepackage{hyperref}       
\usepackage{url}            
\usepackage{booktabs}       
\usepackage{amsfonts}       
\usepackage{subfig}         
\usepackage{nicefrac}       
\usepackage{microtype}      
\usepackage{xcolor}         
\usepackage{graphicx} 
\usepackage{amssymb}
\usepackage{amsmath}
\usepackage{caption}
\usepackage{multirow}
\usepackage{multicol}

\usepackage{subcaption}
\usepackage{algorithm}
\usepackage{algorithmic}
\usepackage{wrapfig}
\title{MTRec: Learning to Align with User Preferences via Mental Reward Models}

\author{%
Mengchen Zhao$^{1}$ \quad Yifan Gao$^{2}$ \quad Yaqing Hou$^2$\thanks{Corresponding author.} \quad Xiangyang Li$^3$ \quad Pengjie Gu$^4$ \\
\textbf{Zhenhua Dong}$^3$ \quad \textbf{Ruiming Tang}$^3$ \quad \textbf{Yi Cai}$^1$\\
$^1$School of Software Engineering, South China University of Technology \\ $^2$School of Computer Science and Technology, Dalian University of Technology \\ $^3$Huawei Noah's Ark Lab \quad $^4$Nanyang Technological University\\
\texttt{\{zzmc, ycai\}@scut.edu.cn}\quad
\texttt{\{otz, houyq\}@mail.dlut.edu.cn}\\
\texttt{\{lixiangyang34, dongzhenhua, tangruiming\}@huawei.com}
}

\begin{document}

\maketitle

\begin{abstract}
Recommendation models are predominantly trained using implicit user feedback, since explicit feedback is often costly to obtain. However, implicit feedback, such as clicks, does not always reflect users' real preferences. For example, a user might click on a news article because of its attractive headline, but end up feeling uncomfortable after reading the content. In the absence of explicit feedback, such erroneous implicit signals may severely mislead recommender systems. In this paper, we propose MTRec, a novel sequential recommendation framework designed to align with real user preferences by uncovering their internal satisfaction on recommended items. Specifically, we introduce a mental reward model to quantify user satisfaction and propose a distributional inverse reinforcement learning approach to learn it. The learned mental reward model is then used to guide recommendation models to better align with users' real preferences. Our experiments show that MTRec brings significant improvements to a variety of recommendation models. We also deploy MTRec on an industrial short video platform and observe a 7\% increase in average user viewing time.
\end{abstract}

\section{Introduction}
In interactive recommender systems, explicit feedback (e.g., ratings) is inherently sparse. Consequently, recommendation models predominantly rely on implicit signals (e.g., clicks) for training. However, such signals frequently fail to capture users’ real preferences. For instance, clicking on a video may not indicate satisfaction with its content, while skipping a video could stem from prior exposure to similar content on other platforms rather than genuine dislike. These observations highlight a fundamental misalignment between recommendation models and users' real preferences.

To mitigate such misalignment caused by erroneous feedback signals, a natural approach is to incentivize users to provide explicit feedback. However, in real-world scenarios, users exhibit low propensity to offer such feedback due to cognitive burdens and interface constraints. Prior studies treat erroneous feedback signals as noisy labels and applying denoising techniques to address them, yet their effectiveness remains limited because erroneous feedback is not random noise by its nature \cite{denoising1}. Some alternative methods attempt to mitigate erroneous feedback via multi-feedback fusion, yet they often struggle when confronted with conflicting feedback \cite{multi_feedback}. Overall, existing works focus on data mining approaches, lacking a deep understanding of the mismatch between users' implicit feedback and their real preferences. 

In this work, we aim to quantify and uncover users' internal satisfaction with recommendations, thereby bridging the gap between the recommendation model and users' real preferences. In fact, each time user takes an action (e.g., consume an item), a private feeling will be generated in her mind, telling how she is satisfied by taking the action. We summarize such private feeling as \textit{mental reward}. We have following two observations on the mental reward. O1: The mental reward will influence user's short-term interests and her subsequent behaviors. For example, if an user clicked on a news but felt uncomfortable with the content, she would lose interests on that topic and probably not click on similar news again. O2: Users are maximizing their accumulated mental rewards. This is reasonable because users naturally pursue good experiences during interaction with the recommender system. The above observations indicate that mental reward plays an important role in user's sequential decision making. If we can directly optimize user's mental rewards, the recommendation model would be better aligned with users' real preferences.

To this end, we propose MTRec, a novel sequential recommendation framework which uses a learned Mental Reward Model to guide the recommendation model to align with users' real preferences. First of all, we model the user's decision making as a Markov Decision Process (MDP). With the assumption that the user always maximizes her accumulated mental rewards, we use Inverse Reinforcement Learning (IRL) to infer a mental reward function from users' behavioral data. However, plain IRL recovers a deterministic mental reward function, which fails to capture the random nature of the mental rewards. To address this, we propose a Quantile Regression Inverse Q-Learning (QR-IQL) approach to learn a distributional mental reward function, which maps a state-action pair to a distribution of mental rewards. Hence, we use the rewards predicted by the mental reward model as complementary supervision signals to guide the training of recommendation model. In such a way, the misalignment between recommendation model and user's real preferences can be greatly reduced. Experiments on two public datasets show that MTRec significantly improves the performance of several popular recommendation models, in terms of Area Under Curve (AUC) and Normalised Capped Importance Sampling (NCIS). We also test MTRec in Virtual Taobao to demonstrate its effectiveness on reinforcement learning based recommendation models. Moreover, we deployed MTRec in a real-world industrial short video recommendation platform and observed a 7\% increase in average user viewing time over a 7-day period during the online A/B test.


Our main contributions are summarized as follows.

\begin{itemize}
    \item We identify the misalignment problem in sequential recommendation, where erroneous user feedback could severely deviate recommendation model from users' real preferences. 

    \item We introduce MTRec, a novel sequential recommendation framework that aims to bridge the gap between the recommendation model and users' real preferences by a learned mental reward model, which uncovers users' internal satisfaction with recommendations.

    \item To capture the random nature of the mental rewards, we develop a distributional variant of IRL called QR-IQL to learn the mental reward model. We show how to use the learned mental reward model to guide the optimization of sequential recommendation models.

    \item We conducted extensive offline and online experiments to demonstrate the improvements brought by MTRec. Additionally, we deployed MTRec in a real-world industrial short video recommendation platform and observed a significant increase in user engagement.

\end{itemize}

\section{Related Works}

\textbf{Implicit user feedback in recommendation.} Since explicit feedback is very sparse, industrial recommender systems rely on implicit feedback (e.g., click, video watching) to train recommendation models \cite{youtube,din,adv_cm,cvr}. However, these models ignore the fact that implicit feedback may not reflect users' real preferences. Some works treat such erroneous feedback as random noise and try to eliminate its influence via denoising, yet these methods have limited accuracy because it is hard to distinguish it from genuine responses \cite{denoising2,denoising1}. Another line of works address various types of biases in user feedback, including position bias, selection bias, popularity bias and exposure bias \cite{bias2,bias1}. However, errors in implicit user feedback differ from the aforementioned biases. In some sense, such errors can be regarded as a certain type of inductive bias \cite{bias3}, because implicit feedback is wrongly assumed to reflect users' real preferences.\\ 


\noindent \textbf{AI alignment through reward model.} Emerging from Natural Language Processing, alignment algorithms have proven effective due to their ability to guide Large Language Models (LLMs) in matching human values. Many popular alignment methods employ a reward model to provide fine-tuning signals. For example, Reinforcement Learning from Human Feedback (RLHF) learns human preferences through a reward model trained with human-rated outputs \cite{rlhf}. Reward rAnked FineTuning (RAFT) uses a reward model to select the best set of training samples based on model outputs \cite{raft}. Inspired by these works, we aim to learn a reward model to guide recommendation models in aligning with users' real preferences. Unlike prior approaches that rely on human annotators to provide reward model labels, our reward model is directly learned from existing user behavioral data, making it more industry-friendly.\\



\noindent \textbf{RL and IRL for sequential recommendation.} Sequential recommendation is typically modeled as interactions between users and recommender systems. Prior works have explored using reinforcement learning (RL) to optimize recommendation policy, where the recommender system is modeled as an agent and the  users are treated as key components of environment \cite{topk,drn}. RL-based methods have great potential to maximize recommender systems' long-term revenue, but they often suffer from the bias of user simulators. Inverse RL (IRL) aims to recover the agent's reward function from expert trajectories. Some works apply IRL to infer the reward function for the recommender system, assuming expert recommendation policies are available \cite{irl4rec1,irl4rec2}. However, these works ignore that the users are also active agents, and understanding user behaviors is essential for improving recommendation policies. In our work, we model users as agents and use IRL to infer the optimal reward model from their behaviors. Although similarly employing IRL techniques, our work differs fundamentally from existing works by inferring a user-centric reward model, as opposed to system-centric reward modeling.


\section{Preliminaries}
\label{seq:pre}
\noindent \textbf{Sequential recommendation.} Typically, a sequential recommendation model takes a sequence of user-item interactions as input and predicts the next items that mostly attract the users. Due to the lack of explicit feedback, sequential recommendation tasks are usually formulated as predicting the next item that is most likely to induce target user behaviors such as clicks:

\begin{equation}
\label{eq:sequential}
  i_t = \arg\max_{i \in I} p_\phi(\hat{a}|i,h_{t-1}), 
\end{equation}

\noindent where $\hat{a}$ represents user's behavior to be predicted, $I=\{i_1,...,i_N\}$ is the set of candidate items. The parameterized function $p_\phi$ measures the probability of the target behavior $\hat{a}$. The interaction history $h_t$ consists of a list of tuples up to time step $t$, where each tuple $<u,i,a>_{t}$ consists of a user $u$, an item $i$ and the user's action $a\in A$. In a simplified case, user's action space $A$ can be restricted to $\{click, skip\}$, hence the recommendation task is reduced to predicting the user's click-trough rate.\\

\noindent \textbf{Inverse reinforcement learning.} Conventionally, maximum entropy IRL aims to learn a deterministic reward function $r:S\times A\rightarrow \mathbb{R}$ by solving Problem~\ref{eq:MaxentIRL} \cite{maxentIRL}. 

\begin{equation}
\label{eq:MaxentIRL}
\max_{r\in\mathcal{R}}\min_{\pi\in\Pi} \mathbb{E}_{\pi_E}[r(s,a)] - \mathbb{E}_{\pi}[r(s,a)] - \mathcal{H}(\pi),
\end{equation}
where $\mathcal{H}(\pi) \triangleq \mathbb{E}_{\pi}[-\log\pi(a|s)]$ denotes the entropy of policy $\pi$. Intuitively, this formulation learns a reward function that assigns high reward to the expert policy $\pi_E$ and a low reward to other policies, while searching for the optimal policy under the reward function in the inner loop. The expectation over policies can be replaced by the occupancy measure $\rho_\pi(s,a) = \pi(a|s)\sum_t\gamma^t P(s_t=s|\pi)$, which specifies a probability distribution over $(s,a)$ pairs. Then, Problem~\ref{eq:MaxentIRL} can rewritten as:

\begin{equation}
\label{eq:IRL-rho}
\max_{r\in\mathcal{R}}\min_{\pi\in\Pi} \mathbb{E}_{\rho_E}[r(s,a)] - \mathbb{E}_{\rho}[r(s,a)] - \mathcal{H}(\pi)- \phi(r),
\end{equation}
where $\phi$ is a convex regularizer on $r$ \cite{gail}.  If we define the regularizer as $\psi(x) = x - \phi(x)$, then $\mathbb{E}_{\rho_E}[r(s,a)] - \phi(r)$ can be compactly represented by $\mathbb{E}_{\rho_E}[\psi(r(s,a))]$. Moreover, 
the reward function $r$ can be represented by the soft $Q$-function as $r(s,a) = Q(s,a) - \gamma \mathbb{E}_{s'\sim P(s,a)}V(s')$, where $V(s) = \mathbb{E}_{a\sim \pi(\cdot|s)}[Q(s,a) - \log \pi(a|s)]$. Also the optimal policy $\pi^*$ can be represented by the soft $Q$-function as $\pi^*(a|s) = \frac{1}{\Delta(s)}\exp(Q(s,a))$, where $\Delta(s)={\sum_{a'}\exp(Q(s,a'))}$ is the normalization factor. Then, the inner problem becomes trivial so that Problem~\ref{eq:IRL-rho} can be reformulated as follows \cite{maxentIRL,sac}.



\begin{equation}
\label{eq:IRL-initial}
\max_{Q\in\Omega} \mathbb{E}_{s,a\sim\rho_E}[\psi(Q(s,a)-\gamma \mathbb{E}_{s'\sim P(s,a)}V^*(s'))]
 - (1-\gamma) \mathbb{E}_{s_0\sim \rho_0}[V^*(s_0)],
\end{equation}

\noindent where $V^*(s)=\log\sum_a\exp Q(s,a)$. Note that the objective of Problem~\ref{eq:IRL-initial} depends only on $Q$, which allows us to solve the problem by directly optimizing a Q-network. Solving Problem~\ref{eq:IRL-initial} with the expert data $D_E$ results in the optimal $Q$-function, and the deterministic rewards can be recovered by:
\begin{equation}
\label{eq:recover_r}
    r(s,a) = Q(s,a) - \gamma \mathbb{E}_{\rho_E}[\log\sum_a\exp Q(s',a)],
\end{equation}

\noindent \textbf{The misalignment problem in recommendation.} In the context of LLM, the alignment problem is defined as optimizing the outputs of LLMs towards matching human values \cite{rlhf}. For recommender systems, the general goal is to maximize users' satisfaction by selecting appropriate items for them. This is naturally in accord with the alignment problem in LLMs, in the sense that aligning with human values is similar to aligning with user preferences. However, directly maximizing user satisfaction is very challenging for recommender systems. Therefore, most existing works focus on optimizing some surrogate objectives such as click-through rate and conversion rate \cite{dcn,dien}. Although the surrogate objectives make it convenient to optimize recommendation models, they suffer from inductive biases since users' implicit feedback may not reflect their real preferences. As a consequence, \textit{the optimization goal of recommendation models may deviate from users' real preferences.} Compared with one-shot recommendation, the misalignment problem in sequential recommendation is even worse due to the accumulation of errors.    

Inspired by recent advances in LLM alignment, we aim to develop a reward model that helps to align recommendation models with users' real preferences. The challenges are two-fold. First, we do not have an off-the-shelf reward model that reflects users' real preferences, under the lack of their explicit feedback. To address this, we propose the Mental Reward model learned from rich user behavioral data to approximate their real preferences. Second, unlike static human values, users' preferences could be highly stochastic, potentially due to unpredictable environmental factors and preference shifts. To resolve this, we propose a distributional Inverse RL approach to capture the randomness of the mental reward model, which will be illustrated in Section~\ref{sec:method}.

\section{Method}
\label{sec:method}
\noindent \textbf{Motivation.} User behaviors have been extensively studied in the literature of recommender systems. However, existing works treat user behaviors as labels for recommendation models. In fact, during the interaction with recommender systems, users are active agents rather than static label providers. Moreover, users are implicitly maximizing their own reward functions by taking actions in recommender systems. Based on the above insights, we believe that uncovering the user's reward function would significantly benefit recommendation models in aligning with users' real preferences.\\

\noindent \textbf{Overview.} We propose a MenTal reward based Recommendation framework MTRec, which consists of three main parts. First, we introduce a novel User-Centric Markov Decision Process, where uses are modeled as active agents during their interaction with recommender systems. Second, we develop a distributional IRL method called Quantile Regression Inverse Q-learning (QR-IQL) for learning the mental reward model. Last, we show how to use the learned mental reward model to guide existing recommendation models to align with users' real preferences.

\begin{figure*}[t]
    \centering
   \includegraphics[width=1\linewidth]{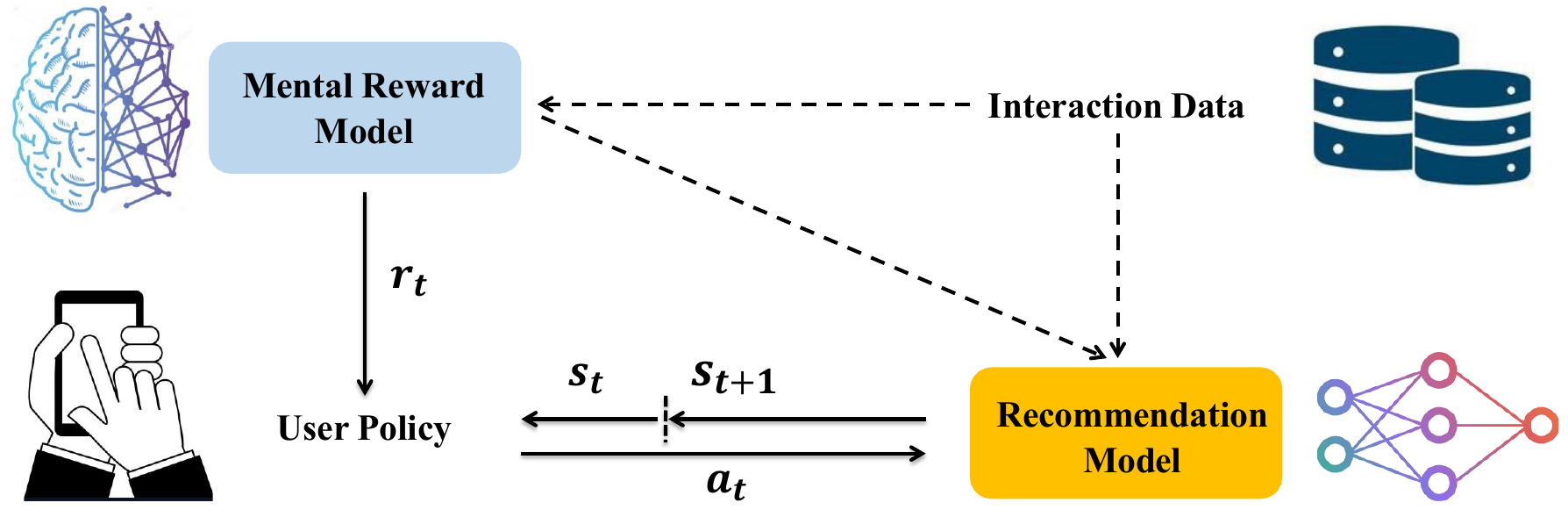}
    \caption{The overall framework of MTRec. The solid lines represent the interaction process. The dashed lines represent the information flow between data and models. Our goal is to recover the mental reward model and use it to improve the recommendation model.}
    \label{fig:framework}
\end{figure*}

\subsection{User-Centric Markov Decision Process}
\label{sec:user mdp}
Reinforcement learning has been applied to model the interaction between users and recommender systems. Existing works usually model the recommender system as agent, who recommends items and receives user feedback as reward \cite{drn,topk}. However, they often ignore the strategic behavior of the users. In this work, we focus on studying users' behaviors, whose decision making can be modeled as a Markov Decision Process $\mathcal{M}=\left \langle S, A, P, R,\pi \right \rangle$:

\begin{itemize}
    \item $S$ denotes the state space. A state $s_t=(h_{t}, i_t)$ includes the interaction history up to time $t$. At each time step, the interaction is recorded by a tuple $<u,i,a>_t$, consisting of an user $u$, displayed item $i$ and the user's action $a$.

    \item $A$ denotes user's action space. An action $a \in \mathcal{A}$ represents user's response to the item displayed to her. User's responses could be of various types, including explicit feedback such as news clicks and video watchings. In this work, we consider a general case where users' responses are either positive or negative, leading to a simplified action space.

    \item $P: S \times A \rightarrow S$ denotes the transition function. Following existing works on sequential recommendation \cite{din,dien,sasrec}, after an user takes an action, a new state $s_{t+1} = s_t \cup \{h_{t+1},i_{t+1}\}$ will be generated. 
    
    \item $R: S\times A \rightarrow \theta(r)$ denotes the mental reward function which maps a state and an action to a distribution of mental reward $r\in\mathbb{R}$. Note that we model the mental reward $r$ as a random variable instead of a deterministic value, since $r$ could be influenced by unpredictable environmental factors and preference shifts. Such a modeling allows us to capture the intrinsic randomness and potentially richer information on user's mental rewards.

    \item $\pi: \mathcal{S}\rightarrow\mu(a)$ denotes the user's behavioral policy, which maps a state to a distribution over actions. A stochastic user policy facilitates reasoning her reward function using IRL.

\end{itemize}

Figure~\ref{fig:framework} illustrates the interaction process between user and recommendation model. At each round, the user receives a recommended item and select an action to respond. After that, a mental reward that summarizes user's satisfaction about the current item is generated. Note that the mental reward is highly correlated with user's real preference, but is unknown to the recommendation model. However, the mental reward will influence the user's subsequent behaviors as she seeks to maximize the accumulated mental rewards. We will focus on estimating the mental rewards in the following parts.

\subsection{Uncovering the Mental Reward Model} \label{sec:irl4recurm}

Inverse reinforcement learning (IRL) aims to find a reward model that explains behaviors of the expert policy \cite{irl}. Since the user naturally maximizes her mental rewards, her policy can be regarded as the expert policy. Therefore, IRL could be used to recover the mental reward model using the interaction data. \textit{However, existing IRL methods focus on deterministic reward models, which fail to capture the intrinsic randomness of the user's mental rewards}. To this end, we will develop a distributional version of IRL algorithm for uncovering the mental reward model.


\subsubsection{A distributional perspective on IRL}
Conventional IRL methods such as MaxEntIRL iteratively optimize the reward function and the policy until convergence \cite{maxentIRL}. Bayesian IRL methods assume that there are multiple reward functions and focus on estimating their posterior distribution using Bayes' rule \cite{bayesianIRL,nonbayesianIRL}. Although Bayesian IRL methods introduce various distributions over reward functions, they are still restricted to deterministic reward functions. By contrast, we aim to learn a distributional reward function where the reward can be stochastic.  

Since we model the user's mental reward $r$ as a random variable, Equation~\ref{eq:recover_r} can be rephrased using a distributional operator as 
    $$\mathcal{T}^\pi r(s,a) :\overset{D}{=} Q(s,a) - \gamma \mathbb{E}_{\rho_E}[\log\sum_a\exp Q(s',a)],$$  
where $X:\overset{D}{=}U$ denotes equality of probability laws, that is, the random variable $X$ is distributed according to the same
law as $U$. As each $Q(s,a)$ function uniquely determines a distribution of $r(s,a)$, learning a distributional reward function is reduced to learning a distributional $Q$-function.

\subsubsection{Quantile Regression Inverse Q-learning (QR-IQL)}

Problem~\ref{eq:IRL-initial} aims to learn a deterministic $Q$-function, while we aim to learn a distributional $Q$-function. Following the QR-DQN \cite{qrdrn}, the distribution of $Q$ can be characterized by a quantile distribution. We denote by $Z$ the variable associated with the distribution of $Q$, that is, $Q(s,a)=\mathbb{E}[Z(s,a)]$. Let $\lambda: S\times A\rightarrow \mathbb{R}^N$ be a parametric model, where $N$ is the number of quantiles. Then a quantile distribution $Z_\lambda$ maps a state-action pair $(s,a)$ to a uniformly probability distribution supported on $\{\lambda_i(s,a)\}_{i=1}^N$. Instead of learning a scalar value $Q(s,a)$, our model will estimate the positions of supports $\{\lambda_i(s,a)\}_{i=1}^N$ and calculate $Q_\lambda(s,a)$ as:
$$Q_\lambda(s,a)=\frac{1}{N}\sum_{i=1}^N\lambda_i(s,a).$$


Recall that in Problem~\ref{eq:IRL-initial}, we optimize a $Q$-network that outputs a scalar $Q$-value. In order to learn the distributional $Q$-function, we change the output layer of the $Q$-network to be of size $|A|\times N$. To derive the objective for the quantile regression inverse Q-learning, we made two modifications on Problem~\ref{eq:IRL-initial}. First, since our mental reward model is learned from offline data, the second term in Problem~\ref{eq:IRL-initial} (i.e., $(1-\gamma) \mathbb{E}_{s_0\sim \rho_0}[V(s_0)]$) can be replaced by $\mathbb{E}_{(s,a)\sim\rho_E}[V(s)-\gamma V(s')]$. In other words, $\mathbb{E}_{s_0\sim\rho_0}[V(s_0)]$ is irrelevant with the initial state distribution $\rho_0$. Second, as is suggested in \cite{iqlearn}, we choose the regularizer as $\psi(x) = x-\frac{1}{4\alpha}x^2$ for the ease of optimization while bounding the rewards. Finally, we formulate the objective of the quantile regression inverse Q-learning as Problem~\ref{eq:quantile_reg}. The complete derivation is provided in Appendix~\ref{ref:derivation_of_P6}.
\begin{equation}
\label{eq:quantile_reg}
\begin{small}
\max_{Q_\lambda} \mathbb{E}_{\rho_E}[Q_\lambda(s,a)-\log \sum_a \exp Q_\lambda(s,a)] \\
- \frac{1}{4\alpha}\mathbb{E}_{\rho_E}[(Q_\lambda(s,a)-\log \sum_a \exp Q_\lambda(s',a))^2],
\end{small}
\end{equation}
We apply the Pinball loss to get the optimal quantile distribution supports $\{\lambda_i(s,a)\}_{i=1}^N$ and the optimal $Q_\lambda^*$, which we can use to calculate the mental rewards as: 
\begin{equation}
r^*(s,a)=Q_\lambda^*(s,a) - \gamma \mathbb{E}_{\rho_E}[\log\sum_a\exp Q_\lambda^*(s',a)].   
\end{equation}
Detailed optimization algorithm is provided in Appendix~\ref{ref:qr-iql}. While building upon ideas from QR-DQN for RL, our QR-IQL is the first distributional IRL algorithm that uncovers the underlying distribution of rewards, which is key to capture the randomness of users' mental rewards.

\subsection{Applications of the Mental Reward model}
\label{sec:align}
Generally, a sequential recommendation model takes user features, candidate item features, interaction histories and some contextual features as input, and output a score for the candidate item used for ranking. For the sake of brevity, we represent the recommendation model as $F_{\zeta}(i_t|h_{t})$, which is parameterized by $\zeta$. Although the mental reward model $r(s,a)$ indicates users' preferences to some extent, it lacks sufficient feature-level modeling and thus cannot be directly used for recommendation. Instead, we use $r(s,a)$ to provide additional learning signals for recommendation models, However, combining $r(s,a)$ with existing recommendation models is non-trivial due to the variety of learning objectives. Fortunately, these objectives fall into several categories. We will use the following two typical examples to illustrate how to use the mental reward model in practice.\\

\noindent\textbf{Classification-based models.} Many sequential recommendation tasks are formulated as binary classification problems with the following Cross Entropy loss:
\begin{small}
\begin{equation}
\label{eq:ce_loss}
    \mathcal{L}_{CE}(\zeta) = -\mathbb{E}_{D_E}[a^P\log(F_{\zeta}(i|h)) + a^N\log(1-F_{\zeta}(i|h))],  
\end{equation}
\end{small}

\noindent where $a^P=1$ and $a^N=0$ indicate user's positive and negative responses, and $F_{\zeta}(i|h)$ represents the estimated probability of clicking on item $i$ based on history $h$. We want the recommendation models to also maximize the expectation of user's mental rewards, leading to the following alignment loss:
\begin{equation}
\mathcal{L}_{Align}(\zeta) = -\mathbb{E}_{D_E} [r^*(s,a^P)\cdot F_{\zeta}(i|h)
  +r^*(s,a^N)\cdot (1-F_{\zeta}(i|h))],
\end{equation}
Then, the final loss for training the recommendation model can be written as a weighted combination of the two losses:
\begin{equation}
\label{eq:f1}
    \mathcal{L}_{Final}(\zeta) = \mathcal{L}_{CE}(\zeta) + \kappa\cdot \mathcal{L}_{Align}(\zeta).
\end{equation}

\noindent\textbf{RL-based models.} In a typical setting, the recommender system is modeled as an agent, who maximizes the accumulated system rewards. We denote by $\hat{r}(h,i,a)$ the system reward after recommending item $i$ and receiving user feedback $a$ given interaction history $h$. In this context, $F_{\zeta}(i|h)$ represents the RL-based recommendation policy, whose goal is to maximize the expectation of accumulated rewards $\hat{r}$:
\begin{equation}
    \mathcal{L}_{RL}(\zeta) = - \mathbb{E}_{i_t\sim F_{\zeta}}[\sum_t\hat{r}(h_{t-1},i_t,a_t)]
\end{equation}
Since we want the recommendation model to also maximize the mental rewards of the user. We simply add the mental reward $r^*$ to $\hat{r}$ and obtain the following objective.

\begin{equation}
\label{eq:f2}
     \mathcal{L}_{Final}(\zeta) = 
      - \mathbb{E}_{i_t\sim F_{\zeta}}[
      \sum_t\hat{r}(h_{t-1},i_t,a_t) \\
       + \kappa\cdot r^*(h_{t-1},i_t,a_t)]
\end{equation}
See Appendix~\ref{app:overall_algorithm} for more implementation details.

\section{Experiments}
\label{sec:exp}
In this section, we report the performance of MTRec in both offline and online settings, with focuses on answering the following research questions (RQs).

\begin{itemize}
    \item (RQ1:) How does MTRec improve classification-based recommendation models?
    
    \item (RQ2:) How does MTRec improve RL-based recommendation models?
    \item (RQ3:) Does the learned mental reward model provide useful information? 
    \item (RQ4:) How does MTRec perform in online A/B test?
\end{itemize}

\subsection{Experiments on Public Datasets (RQ1)}
\label{sec:offline_exp}

\noindent \textbf{Datasets.} The Amazon dataset \cite{amazon} collects user review data from Amazon e-commerce platform. We use two subsets of the Amazon dataset: Books and Electronics in our offline experiments. More details on processing the datasets are provided in Appendix~\ref{seq:stats_amazon}.


\noindent \textbf{Baselines.}
We use eight widely used recommendation models as baselines and combine each of them with MTRec to test the improvements brought by MTRec. \textbf{Wide\&Deep} \cite{widedeep} is a hybrid recommendation model combining a wide linear model and deep neural network for collaborative filtering. \textbf{PNN} \cite{pnn} is a neural network architecture designed for CTR prediction in recommender systems. \textbf{DeepFM} \cite{deepfm} is a hybrid recommendation model combining factorization machines and deep neural networks. \textbf{SASRec} \cite{sasrec} uses the self-attention mechanism to model sequential patterns for recommendation systems. \textbf{DIN} \cite{din} is an attention-based neural model for sequential recommendation, where the attention mechanism aims to distinguish the interest of a user's historical behaviors. \textbf{DIEN} \cite{dien} designs a sequential architecture to model interest evolution for recommendation, which uses an auxiliary loss to capture temporal interests. \textbf{LinRec} \cite{linrec} is a lightweight linear recommendation model designed for efficient computation and scalability with large datasets. \textbf{SIGMA} \cite{sigma} is a sequential recommendation model that uses a selective gating mechanism to focus on the most relevant user behaviors for improved performance.\\

\noindent \textbf{Evaluation metrics.}
We use the following two metrics: Area Under Curve (AUC) \cite{auc} and Normalised Capped Importance Sampling (NCIS) \cite{ncis}. AUC is used to measure the model's ranking ability and NCIS is used to approximate the model's online performance \cite{ncis2}. Formally, the score of NCIS can be calculated by:
$$\Tilde{J}^{NCIS}(\mathcal{M}) = \frac{\sum\nolimits_{i=1}^n \Tilde{\rho_i}(\mathcal{M}) * L_i}{\sum\nolimits_{i=1}^n \Tilde{\rho_i}(\mathcal{M})},$$

where $\rho_i(\mathcal{M})=\prod_{\substack{t\in \mathcal{T}}} p_t(\mathcal{M})$ is the probability that the CTR model $\mathcal{M}$ follows the request trajectory of the user $i$, $p_t(\mathcal{M})$ is the click-through rate estimate of model $\mathcal{M}$ for item $t$ and $n$ is the number of users in the test set for NCIS. 
In the experiments, we use the complete trajectories of 10\% of users to calculate NCIS. Moreover,  we obtain the final NCIS score by substracting the NCIS of the untrained model from that of the trained model to eliminate the impact of random parameters among different models.
Intuitively, $\Tilde{J}^{NCIS}(\mathcal{M})$ awards a CTR model with a high score if the model has large probability to follow long trajectories.\\


\begin{wraptable}{r}{0.5\textwidth}
    \vspace{-0.4cm}
    \centering
    \resizebox{0.46\textwidth}{!}{
    \begin{tabular}[t]{lrrrr}
    \toprule
    \multirow{2}{*}{Model} & \multicolumn{2}{c}{Electronics} & \multicolumn{2}{c}{Books}\\
    \cline{2-5}
    & AUC & NCIS  & AUC & NCIS\\
    \hline
        Wide\&Deep & 0.8290 & 0.6749 & 0.8605 & 2.0967\\ 
        Wide\&Deep-IRL & 0.8342 & 0.8738 & \textbf{0.8661} & 3.1264\\
        Wide\&Deep-MTRec & \textbf{0.8351} & \textbf{0.9063} & 0.8657 & \textbf{3.2043}\\
    \hline
        PNN & 0.8396 & 0.6618 & 0.8603 & 2.3316 \\ 
        PNN-IRL & \textbf{0.8547} & 0.9297 & 0.8673 & 3.4567 \\
        PNN-MTRec & 0.8542 & \textbf{0.9515} & \textbf{0.8679} & \textbf{3.4971}\\
    \hline
        DeepFM & 0.8424 & 0.6993 & 0.8634 & 2.3146 \\ 
        DeepFM-IRL & 0.8458 & 0.8542 & 0.8715 & 3.7296 \\
        DeepFM-MTRec & \textbf{0.8468} & \textbf{0.8961} & \textbf{0.8742} & \textbf{3.7904}\\
    \hline
        SASRec & 0.8325 & 0.8243 & 0.8675 & 3.1755\\ 
        SASRec-IRL & \textbf{0.8366} & 0.8681 & 0.8681 & 3.1946\\
        SASRec-MTRec & 0.8328 & \textbf{0.8833} & \textbf{0.8798} & \textbf{3.4634}\\
    \hline
        DIN & 0.8523 & 0.6044 & 0.8653 & 2.1324\\ 
        DIN-IRL & 0.8533 & 0.8168 & 0.8701 & 3.1675 \\
        DIN-MTRec & \textbf{0.8542} & \textbf{0.8728} & \textbf{0.8732} & \textbf{3.2208}\\
    \hline
        DIEN & 0.8448 & 0.7766 & 0.8686 & 2.2685\\ 
        DIEN-IRL & 0.8461 & 0.8857 & 0.8723 &  3.0476 \\
        DIEN-MTRec & \textbf{0.8472} & \textbf{0.9324} & \textbf{0.8757} & \textbf{3.1185}\\
    \hline
        LinRec & 0.8579 & 0.8077 & 0.8754 & 2.4653\\ 
        LinRec-IRL & \textbf{0.8597} & 0.9365 & 0.8771 &  \textbf{3.8005} \\
        LinRec-MTRec & 0.8594 & \textbf{0.9782} & \textbf{0.8792} & 3.7928\\
    \hline
        SIGMA & 0.8581 & 0.7946 & 0.8762 & 2.3975\\ 
        SIGMA-IRL & 0.8592 & 0.9261 & 0.8802 & 3.7556 \\
        SIGMA-MTRec & \textbf{0.8604} & \textbf{0.9563} & \textbf{0.8814} & \textbf{3.8025}\\
    \bottomrule
    \end{tabular}
    }
    \captionsetup{font=small}
    \caption{Experimental results on Amazon datasets.}
    
    \label{table:results on public datasets}
\end{wraptable}
\noindent\textbf{Results on Amazon datasets.} Table \ref{table:results on public datasets} summarizes the experimental results on the two Amazon datasets. The suffix ``IRL" indicates that the mental reward model is learned by a non-distributional IRL algorithm IQ-Learn \cite{iqlearn}, while the suffix ``MTRec" indicates that the mental reward model is learned by our algorithm QR-IQL. It can be observed that integrating MTRec with existing models consistently improves their AUC and NCIS across almost all baseline models. Moreover, the improvements on NCIS are generally more significant than AUC. This aligns with the motivation of MTRec, that is, to maximize overall user satisfaction and long-term engagement. Note that the AUC is also slightly improved, demonstrating that the improvement of users' long-term engagement is not at the sacrifice of the model's ranking ability. In addition, the comparisons between IRL and MTRec demonstrates the benefit of learning a distributional version of mental reward model.

\subsection{Experiments on Virtual Taobao (RQ2)}

In this set of experiments, we choose RL-based recommendation models as our baselines and combine them with MTRec to test their performance. Both training and testing of the algorithms are conducted in simulated interactive recommendation environments on Virtual Taobao \cite{virtualtb}. We construct an expert dataset containing 100,000 high-quality trajectories by recording trajectories with averaged CTR$>$0.5. After training the mental reward model using the expert data, we add the predicted mental rewards to the original rewards and train the baselines again. \\


\noindent\textbf{Baselines.} Virtual Taobao allows training recommendation policies by RL. We use Proximal Policy Optimization (PPO) \cite{ppo} and Soft Actor-Critic (SAC) \cite{sac} as  baselines. According to Equation~\ref{eq:f2}, RL-based recommendation models can be adjusted by simply adding the mental rewards to the base rewards $r_{final}=r_{env} + \kappa\cdot r_{mental}$. In Virtual Taobao, $r_{env}$ is provided by a pre-trained user model. We set $\kappa=0.2$ for trade-off between the two reward signals.\\

\noindent\textbf{Evaluation metric.}
The major evaluation metric used in Virtual Taobao is the episodic Click-Through-Rate (eCTR) during the simulated online interaction. The eCTR is calculated as: $$\text{eCTR}=\frac{r_{episode}}{10*N_{step}},$$
where 10 is the number of items recommended in a single page, $r_{episode}$ is the total number of clicks in an episode and $N_{step}$ is the total number of steps.\\


\begin{figure}
    \centering
    \subfloat{\includegraphics[scale=0.3]{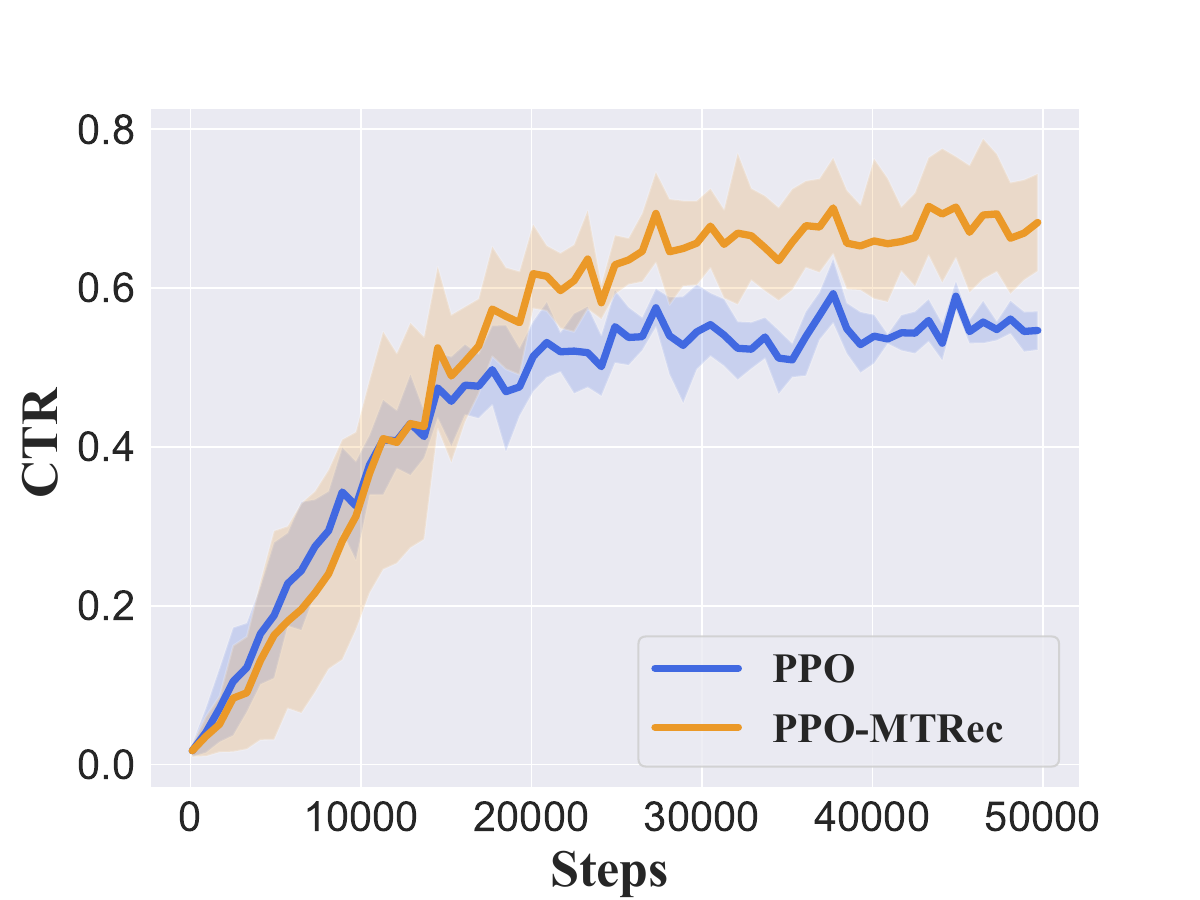}}
    \subfloat{\includegraphics[scale=0.3]{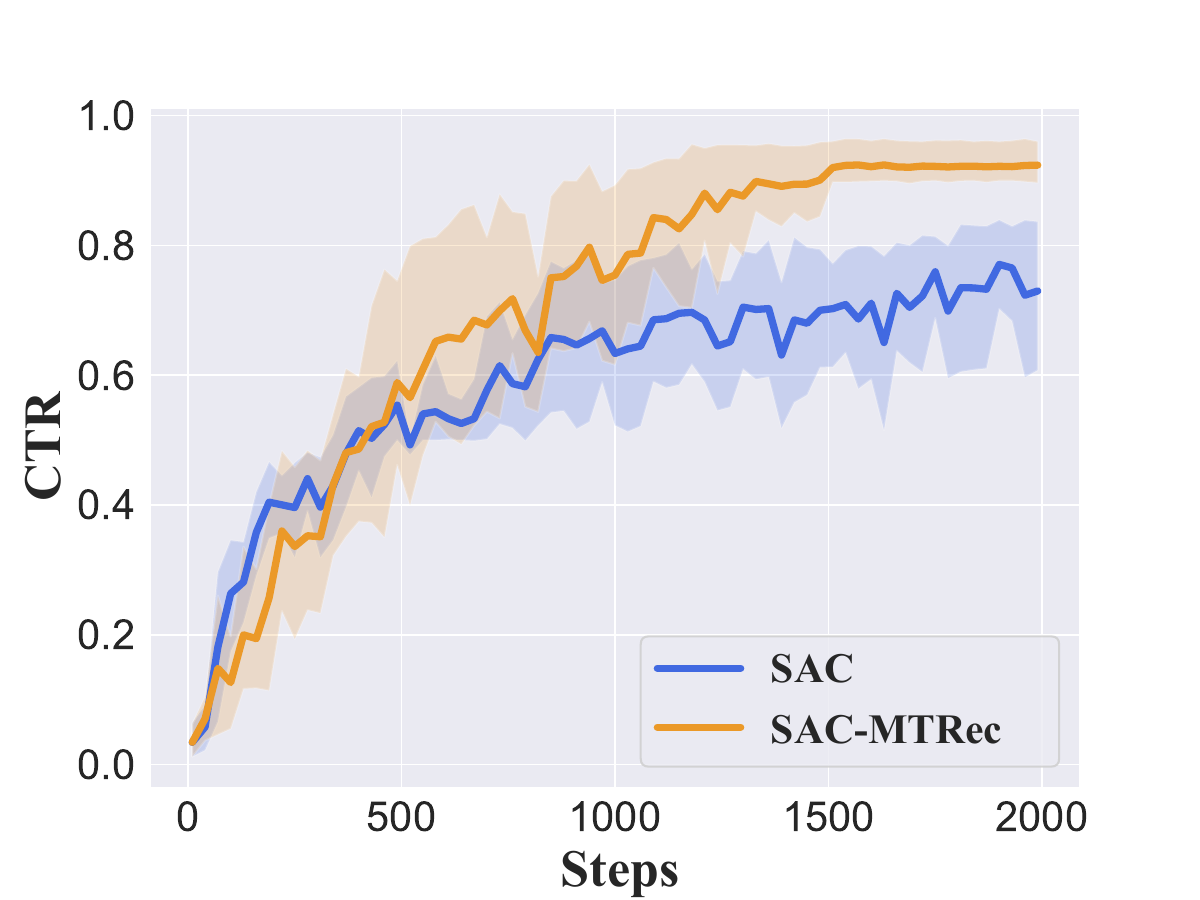}}
    \captionsetup{font=small}
    \caption{Training curves of RL models. Averaged CTR is reported with 95\% confidence interval.}
    \label{fig:vt_result}
\end{figure}
\noindent \textbf{Results on Virtual Taobao.} From Fig \ref{fig:vt_result} we can see that the baselines attain average CTRs of 0.5435 and 0.7055 respectively. By incorporating the mental rewards during training, PPO and SAC show significantly improved performance at average CTRs of 0.678 and 0.909 respectively. This demonstrates that the mental rewards provide more useful information about users' real preferences and successfully boost the performance of RL-based recommendation models.

\begin{figure}
    \centering
   \includegraphics[width=1\linewidth]{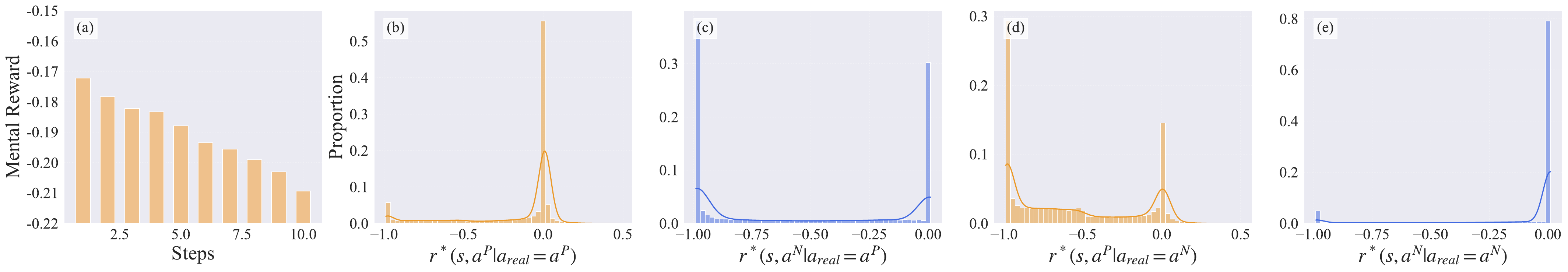}
   \captionsetup{font=small}
    \caption{Illustrations of the predicted mental rewards. (a) Averaged mental rewards by steps in all trajectories; (b-e) Expected and counterfactual mental rewards given actual user actions.}

    \label{fig:illustration}
\end{figure}

\subsection{Evaluation of the Mental Reward Model (RQ3)}
\label{sec:eval_mtmodel}
It is almost impossible to evaluate the learned mental reward model directly due to the lack of ground-truth mental reward labels. Yet, we evaluate the mental reward model indirectly by its correlation with the datasets. First, we calculate the averaged mental rewards at different steps in all trajectories. Figure~\ref{fig:illustration} (a) shows the results of the first ten steps on Amazon Book dataset. We can see that the averaged mental rewards decrease obviously with the increase of steps. This result aligns with our intuition that users may get tired and receive less mental rewards during their interaction with the recommender system. 

Second, we visualize the distribution of the mental rewards conditioned on the users' real responses. For instance, Figure~\ref{fig:illustration} (b) and (c) show the distributions of expected mental reward $r^*(s,a^P|a_{real}=a^P)$ and the counterfactual mental reward $r^*(s,a^N|a_{real}=a^P)$ when the users take a positive action $a_{real}=a^P$. Specifically, $r^*(s,a^P|a_{real}=a^P)$ represents the predicted user's mental reward after she takes action $a^P$, while $r^*(s,a^N|a_{real}=a^P)$ represents the counterfactual mental reward if she had taken a negative action $a^N$. Intuitively, an user should receive a relatively high reward for the action she actually taken (a high $r^*(s,a_{real}|a_{real})$). Yet, the experimental results show that a non-negligible proportion of $r^*(s,a_{real}|a_{real})$ is relatively low, which suggests that there is a mismatch between users' actions and their real preferences (recall the example in the abstract: a user might click on a news article because of its attractive headline, but end up feeling uncomfortable after reading it). These observations indirectly validate the effectiveness of the mental reward model.

\subsection{Online A/B test on Industrial Platform (RQ4)}
To further validate the effectiveness of MTRec, we deploy it on an industrial short-video recommendation platform, which has tens of millions of Daily Active Users (DAU). Short-video recommendation is a typical sequential recommendation scenario, where the goal is to improve the user engagement with the platform. The baseline model is a Deep Cross Network (DCN) \cite{dcn}, which is trained using binary labels indicating whether the users click on videos. However, intuitively, clicking on a video does not necessarily mean that the user is satisfied after watching the video. Therefore, we expect that MTRec could help to improve the overall users' satisfaction on the recommended videos and hence improve their engagement.\\

\begin{wrapfigure}{r}{6cm}
\vspace{-0.9cm}
    \centering
   \includegraphics[width=1\linewidth]{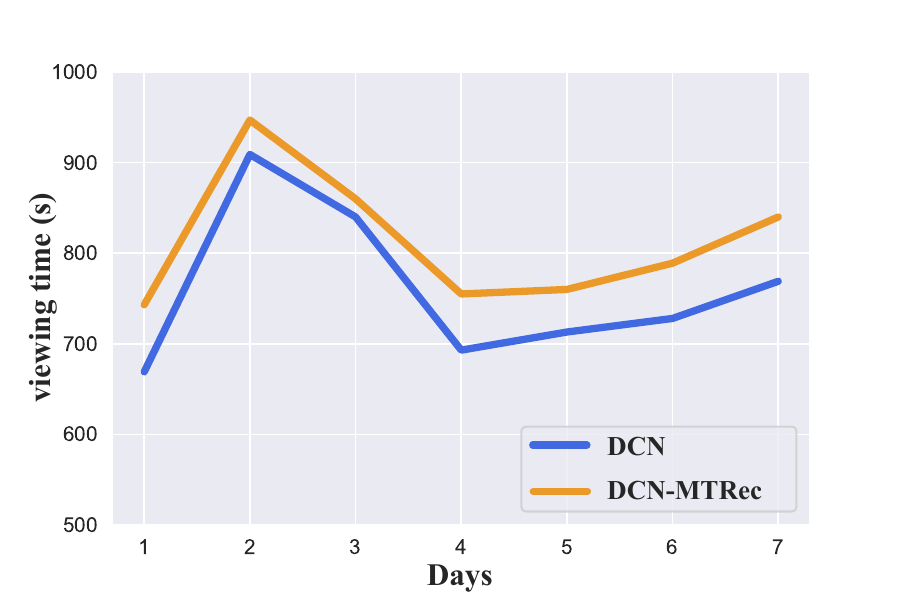}
   \captionsetup{font=small}
    \caption{Online A/B test results.}
    \label{fig:online}
\end{wrapfigure}
\label{sec:conclude_limitation}
We improve the DCN model by incorporating the alignment loss, as illustrated in Equation \ref{eq:f1}. We find that the averaged video viewing time stably improve by about 7\% compared with the baseline model, which demonstrates that MTRec indeed improves the overall recommendation quality and leads to better user engagement. Note that MTRec is quite industrial friendly because of two reasons: (1) the dataset used to train MTRec could be same with that used to train the recommendation model; (2) we only need to add an auxiliary alignment loss to the original recommendation loss.

\section{Conclusions}
The general goal of recommender systems is to satisfy users by providing items that align with their real preferences. However, existing works focus on optimizing surrogate objectives based on users' implicit feedback, ignoring that the implicit feedback may not accurately reflect their real preferences. Consequently, recommendation models could be systematically biased. In this work, we aim to fill this gap by studying the users' behaviors and uncovering the distribution of their mental rewards. We propose a novel distributional IRL algorithm to learn the mental reward model and use it to guide the training of recommendation models. Finally, we validate the effectiveness of MTRec via both offline and online experiments, including the A/B on an industrial short video recommendation platform. In the end, we believe that existing studies on user modeling are far from totally understanding users. 


\section*{Acknowledgements}

This research is supported by Guangdong Basic and Applied Basic Research Foundation (2025A1515010247), and the Fundamental Research Funds for the Central Universities (2024ZYGXZR069).

\newpage
\bibliography{reference}

\begin{thebibliography}{39}
\providecommand{\natexlab}[1]{#1}
\providecommand{\url}[1]{\texttt{#1}}
\expandafter\ifx\csname urlstyle\endcsname\relax
  \providecommand{\doi}[1]{doi: #1}\else
  \providecommand{\doi}{doi: \begingroup \urlstyle{rm}\Url}\fi

\bibitem[Brian et~al.(2008)Brian, Maas, Bagnell, and Dey]{maxentIRL}
Ziebart Brian, Andrew Maas, Andrew Bagnell, and Anind Dey.
\newblock {Maximum entropy inverse reinforcement learning}.
\newblock In \emph{Proceedings of the 23rd AAAI Conference on Artificial Intelligence}, pages 1433--1438, 2008.

\bibitem[Chen et~al.(2021{\natexlab{a}})Chen, Chen, Wang, Xie, Wang, Xia, and Zhu]{multi_feedback}
Hong Chen, Yudong Chen, Xin Wang, Ruobing Xie, Rui Wang, Feng Xia, and Wenwu Zhu.
\newblock Curriculum disentangled recommendation with noisy multi-feedback.
\newblock In \emph{Advances in Neural Information Processing Systems}, pages 26924--26936, 2021{\natexlab{a}}.

\bibitem[Chen et~al.(2019)Chen, Beutel, Covington, Jain, Belletti, and Chi]{topk}
Minmin Chen, Alex Beutel, Paul Covington, Sagar Jain, Francois Belletti, and Ed~H. Chi.
\newblock {Top-k off-policy correction for a REINFORCE recommender system}.
\newblock In \emph{Proceedings of the Twelfth ACM International Conference on Web Search and Data Mining}, pages 456--464, 2019.

\bibitem[Chen et~al.(2021{\natexlab{b}})Chen, Yao, Sun, Wang, Xu, and Zhu]{irl4rec1}
Xiaocong Chen, Lina Yao, Aixin Sun, Xianzhi Wang, Xiwei Xu, and Liming Zhu.
\newblock Generative inverse deep reinforcement learning for online recommendation.
\newblock In \emph{Proceedings of the 30th ACM International Conference on Information and Knowledge Management}, pages 201--210, 2021{\natexlab{b}}.

\bibitem[Cheng et~al.(2016)Cheng, Koc, Harmsen, Shaked, Chandra, Aradhye, Anderson, Corrado, Chai, Ispir, et~al.]{widedeep}
Heng-Tze Cheng, Levent Koc, Jeremiah Harmsen, Tal Shaked, Tushar Chandra, Hrishi Aradhye, Glen Anderson, Greg Corrado, Wei Chai, Mustafa Ispir, et~al.
\newblock Wide \& deep learning for recommender systems.
\newblock In \emph{Proceedings of the 1st workshop on deep learning for recommender systems}, pages 7--10, 2016.

\bibitem[Covington et~al.(2016)Covington, Adams, Sargin, and Zhao]{youtube}
Paul Covington, Jay Adams, Emre Sargin, and Mengchen Zhao.
\newblock Deep neural networks for youtube recommendations.
\newblock In \emph{Proceedings of the 10th ACM Conference on Recommender Systems}, page 191–198, 2016.

\bibitem[Dai et~al.(2021)Dai, Lin, Zhang, Li, Liu, Tang, He, Hao, Wang, and Yu]{adv_cm}
Xinyi Dai, Jianghao Lin, Weinan Zhang, Shuai Li, Weiwen Liu, Ruiming Tang, Xiuqiang He, Jianye Hao, Jun Wang, and Yong Yu.
\newblock {An adversarial imitation click model for information retrieval}.
\newblock In \emph{Proceedings of the Web Conference}, pages 1809--1820, 2021.

\bibitem[Deepak and Amir(2007)]{bayesianIRL}
Ramachandran Deepak and Eyal Amir.
\newblock Bayesian inverse reinforcement learning.
\newblock In \emph{Proceedings of the 20th International Joint Conference on Artificial Intelligence}, pages 2586--2591, 2007.

\bibitem[Dong et~al.(2023)Dong, Xiong, Goyal, Pan, Diao, Zhang, Shum, and Zhang]{raft}
Hanze Dong, Wei Xiong, Deepanshu Goyal, Rui Pan, Shizhe Diao, Jipeng Zhang, Kashun Shum, and Tong Zhang.
\newblock Raft: Reward ranked finetuning for generative foundation model alignment, 2023.

\bibitem[Fawcett(2006)]{auc}
Tom Fawcett.
\newblock An introduction to roc analysis.
\newblock \emph{Pattern recognition letters}, 27\penalty0 (8):\penalty0 861--874, 2006.

\bibitem[Garg et~al.(2021)Garg, Chakraborty, Cundy, Song, and Ermon]{iqlearn}
Divyansh Garg, Shuvam Chakraborty, Chris Cundy, Jiaming Song, and Stefano Ermon.
\newblock Iq-learn: Inverse soft-q learning for imitation.
\newblock In \emph{Advances in Neural Information Processing Systems}, pages 4028--4039, 2021.

\bibitem[Gilotte et~al.(2018)Gilotte, Calauz{\`e}nes, Nedelec, Abraham, and Doll{\'e}]{ncis2}
Alexandre Gilotte, Cl{\'e}ment Calauz{\`e}nes, Thomas Nedelec, Alexandre Abraham, and Simon Doll{\'e}.
\newblock Offline a/b testing for recommender systems.
\newblock In \emph{Proceedings of the Eleventh ACM International Conference on Web Search and Data Mining}, pages 198--206, 2018.

\bibitem[Guo et~al.(2017)Guo, Tang, Ye, Li, and He]{deepfm}
Huifeng Guo, Ruiming Tang, Yunming Ye, Zhenguo Li, and Xiuqiang He.
\newblock Deepfm: a factorization-machine based neural network for ctr prediction.
\newblock In \emph{Proceedings of the 26th International Joint Conference on Artificial Intelligence}, 2017.

\bibitem[Haarnoja et~al.(2018)Haarnoja, Zhou, Abbeel, and Levine]{sac}
Tuomas Haarnoja, Aurick Zhou, Pieter Abbeel, and Sergey Levine.
\newblock Soft actor-critic: Off-policy maximum entropy deep reinforcement learning with a stochastic actor.
\newblock In \emph{International conference on machine learning}, pages 1861--1870. PMLR, 2018.

\bibitem[Jaedeug and Kim(2012)]{nonbayesianIRL}
Choi Jaedeug and Kee-Eung Kim.
\newblock Nonparametric bayesian inverse reinforcement learning for multiple reward functions.
\newblock In \emph{Advances in Neural Information Processing Systems}, pages 305--313, 2012.

\bibitem[Jiawei et~al.(2023)Jiawei, Dong, Wang, Feng, Wang, and He]{bias3}
Chen Jiawei, Hande Dong, Xiang Wang, Fuli Feng, Meng Wang, and Xiangnan He.
\newblock Bias and debias in recommender system: A survey and future directions.
\newblock \emph{ACM Transactions on Information Systems}, 41\penalty0 (3):\penalty0 1--39, 2023.

\bibitem[Jonathan and Ermon(2016)]{gail}
Ho~Jonathan and Stefano Ermon.
\newblock Generative adversarial imitation learning.
\newblock In \emph{Advances in Neural Information Processing Systems}, 2016.

\bibitem[Kingma and Ba(2015)]{adam}
Diederik~P. Kingma and Jimmy Ba.
\newblock Deepfm: a factorization-machine based neural network for ctr prediction.
\newblock In \emph{International Conference on Learning Representations}, 2015.

\bibitem[Liu et~al.(2023{\natexlab{a}})Liu, Su, He, Zhao, and Li]{irl4rec2}
Jialin Liu, Xinyan Su, Zeyu He, Xiangyu Zhao, and Jun Li.
\newblock Adversarial batch inverse reinforcement learning: Learn to reward from imperfect demonstration for interactive recommendation.
\newblock In \emph{arXiv preprint arXiv:2310.19536}, 2023{\natexlab{a}}.

\bibitem[Liu et~al.(2023{\natexlab{b}})Liu, Cai, Zhang, Zhao, Gao, Wang, Lv, Fan, Wang, He, et~al.]{linrec}
Langming Liu, Liu Cai, Chi Zhang, Xiangyu Zhao, Jingtong Gao, Wanyu Wang, Yifu Lv, Wenqi Fan, Yiqi Wang, Ming He, et~al.
\newblock Linrec: Linear attention mechanism for long-term sequential recommender systems.
\newblock In \emph{Proceedings of the 46th International ACM SIGIR Conference on Research and Development in Information Retrieval}, pages 289--299, 2023{\natexlab{b}}.

\bibitem[Liu et~al.(2024)Liu, Liu, Wang, Wang, Jia, Wang, Liu, Chang, and Zhao]{sigma}
Ziwei Liu, Qidong Liu, Yejing Wang, Wanyu Wang, Pengyue Jia, Maolin Wang, Zitao Liu, Yi~Chang, and Xiangyu Zhao.
\newblock Bidirectional gated mamba for sequential recommendation.
\newblock \emph{arXiv preprint arXiv:2408.11451}, 2024.

\bibitem[McAuley et~al.(2015)McAuley, Targett, Shi, and Van Den~Hengel]{amazon}
Julian McAuley, Christopher Targett, Qinfeng Shi, and Anton Van Den~Hengel.
\newblock Image-based recommendations on styles and substitutes.
\newblock In \emph{Proceedings of the 38th international ACM SIGIR conference on research and development in information retrieval}, pages 43--52, 2015.

\bibitem[Ouyang et~al.(2022)Ouyang, Wu, Jiang, Almeida, Wainwright, Mishkin, Zhang, Agarwal, Slama, Gray, Schulman, Hilton, Kelton, Miller, Simens, Askell, Welinder, Christiano, Leike, and Lowe]{rlhf}
Long Ouyang, Jeffrey Wu, Xu~Jiang, Diogo Almeida, Carroll Wainwright, Pamela Mishkin, Chong Zhang, Sandhini Agarwal, Katarina Slama, Alex Gray, John Schulman, Jacob Hilton, Fraser Kelton, Luke Miller, Maddie Simens, Amanda Askell, Peter Welinder, Paul Christiano, Jan Leike, and Ryan Lowe.
\newblock Training language models to follow instructions with human feedback.
\newblock In \emph{Advances in Neural Information Processing Systems}, pages 27730--27744, 2022.

\bibitem[Qu et~al.(2016)Qu, Cai, Ren, Zhang, Yu, Wen, and Wang]{pnn}
Yanru Qu, Han Cai, Kan Ren, Weinan Zhang, Yong Yu, Ying Wen, and Jun Wang.
\newblock Product-based neural networks for user response prediction.
\newblock In \emph{2016 IEEE 16th international conference on data mining (ICDM)}, pages 1149--1154. IEEE, 2016.

\bibitem[Saurabh and Doshi(2021)]{irl}
Arora Saurabh and Prashant Doshi.
\newblock A survey of inverse reinforcement learning: Challenges, methods and progress.
\newblock \emph{Artificial Intelligence}, 297\penalty0 (103500), 2021.

\bibitem[Schulman et~al.(2017)Schulman, Wolski, Dhariwal, Radford, and Klimov]{ppo}
John Schulman, Filip Wolski, Prafulla Dhariwal, Alec Radford, and Oleg Klimov.
\newblock Proximal policy optimization algorithms.
\newblock \emph{arXiv preprint arXiv:1707.06347}, 2017.

\bibitem[Shi et~al.(2019)Shi, Yu, Da, Chen, and Zeng]{virtualtb}
Jing-Cheng Shi, Yang Yu, Qing Da, Shi-Yong Chen, and An-Xiang Zeng.
\newblock Virtual-taobao: Virtualizing real-world online retail environment for reinforcement learning.
\newblock In \emph{Proceedings of the AAAI Conference on Artificial Intelligence}, volume~33, pages 4902--4909, 2019.

\bibitem[Swaminathan and Joachims(2015)]{ncis}
Adith Swaminathan and Thorsten Joachims.
\newblock The self-normalized estimator for counterfactual learning.
\newblock \emph{advances in neural information processing systems}, 28, 2015.

\bibitem[Wang et~al.(2017)Wang, Fu, Fu, and Wang]{dcn}
Ruoxi Wang, Bin Fu, Gang Fu, and Mingliang Wang.
\newblock Deep \& cross network for ad click predictions.
\newblock In \emph{Proceedings of the ADKDD'17}, 2017.

\bibitem[Wang et~al.(2021)Wang, Feng, He, Nie, and Chua]{denoising1}
Wenjie Wang, Fuli Feng, Xiangnan He, Liqiang Nie, and Tat-Seng Chua.
\newblock Denoising implicit feedback for recommendation.
\newblock In \emph{Proceedings of the 14th ACM International Conference on Web Search and Data Mining}, pages 373--381, 2021.

\bibitem[Wang-Cheng and McAuley(2018)]{sasrec}
Kang Wang-Cheng and Julian McAuley.
\newblock Self-attentive sequential recommendation.
\newblock In \emph{Proceedings of the International Conference on Data Mining}, pages 197--206, 2018.

\bibitem[Will et~al.(2018)Will, Rowland, Bellemare, and Munos]{qrdrn}
Dabney Will, Mark Rowland, Marc Bellemare, and Rémi Munos.
\newblock {Distributional reinforcement learning with quantile regression}.
\newblock In \emph{Proceedings of the 33rd AAAI Conference on Artificial Intelligence}, 2018.

\bibitem[woong Lee et~al.(2021)woong Lee, Park, Lee, and mengchen zhao]{bias2}
Jae woong Lee, Seongmin Park, Jongwuk Lee, and mengchen zhao.
\newblock Dual unbiased recommender learning for implicit feedback.
\newblock In \emph{Proceedings of the 44th International ACM SIGIR Conference on Research and Development in Information Retrieval}, pages 1647--1651, 2021.

\bibitem[Yang et~al.(2021)Yang, Li, Han, Zhuang, Zhan, Zeng, and Tong]{cvr}
Jiaqi Yang, Xiang Li, Shuguang Han, Tao Zhuang, Dechuan Zhan, Xiaoyi Zeng, and Bin Tong.
\newblock {Capturing delayed feedback in conversion rate prediction via elapsed-time sampling}.
\newblock In \emph{Proceedings of the AAAI Conference on Artificial Intelligence}, pages 4582--4589, 2021.

\bibitem[Yi et~al.(2023)Yi, Tang, Rong, and Zhu]{bias1}
Ren Yi, Hongyan Tang, Jiangpeng Rong, and Siwen Zhu.
\newblock Unbiased pairwise learning from implicit feedback for recommender systems without biased variance control.
\newblock In \emph{Proceedings of the 46th International ACM SIGIR Conference on Research and Development in Information Retrieval}, pages 2461--2465, 2023.

\bibitem[Yu and Qin.(2020)]{denoising2}
Wenhui Yu and Zheng Qin.
\newblock Sampler design for implicit feedback data by noisy-label robust learning.
\newblock In \emph{Proceedings of the 43rd International ACM SIGIR Conference on Research and Development in Information Retrieval}, pages 861--870, 2020.

\bibitem[Zheng et~al.(2018)Zheng, Zhang, Zheng, Xiang, Yuan, Xie, and Li]{drn}
Guanjie Zheng, Fuzheng Zhang, Zihan Zheng, Yang Xiang, Nicholas~Jing Yuan, Xing Xie, and Zhenhui Li.
\newblock {DRN: A deep reinforcement learning framework for news recommendation}.
\newblock In \emph{Proceedings of the 2018 world wide web conference}, pages 167--176, 2018.

\bibitem[Zhou et~al.(2018)Zhou, Zhu, Song, Fan, Zhu, Ma, Yan, Jin, Li, and Gai]{din}
Guorui Zhou, Xiaoqiang Zhu, Chenru Song, Ying Fan, Han Zhu, Xiao Ma, Yanghui Yan, Junqi Jin, Han Li, and Kun Gai.
\newblock Deep interest network for click-through rate prediction.
\newblock In \emph{Proceedings of the 24th ACM SIGKDD international conference on knowledge discovery and data mining}, pages 1059--1068, 2018.

\bibitem[Zhou et~al.(2019)Zhou, Mou, Fan, Pi, Bian, Zhou, Zhu, and Gai]{dien}
Guorui Zhou, Na~Mou, Ying Fan, Qi~Pi, Weijie Bian, Chang Zhou, Xiaoqiang Zhu, and Kun Gai.
\newblock Deep interest evolution network for click-through rate prediction.
\newblock In \emph{Proceedings of the AAAI conference on artificial intelligence}, volume~33, pages 5941--5948, 2019.

\end{thebibliography}
\bibliographystyle{plainnat}

\section*{NeurIPS Paper Checklist}

\begin{enumerate}

\item {\bf Claims}
    \item[] Question: Do the main claims made in the abstract and introduction accurately reflect the paper's contributions and scope?
    \item[] Answer: \answerYes{} 
    \item[] Justification: Our main contribution is the MTRec framework with the QR-IQL algorithm, which is described in Section~\ref{sec:method} and justified in Section~\ref{sec:exp}.

\item {\bf Limitations}
    \item[] Question: Does the paper discuss the limitations of the work performed by the authors?
    \item[] Answer: \answerYes{} 
    \item[] Justification: The limitation is discussed in Appendix~\ref{sec:limitation}.

\item {\bf Theory assumptions and proofs}
    \item[] Question: For each theoretical result, does the paper provide the full set of assumptions and a complete (and correct) proof?
    \item[] Answer: \answerYes{} 
    \item[] Justification: The derivations details are provided in Appendix~\ref{sec:appendix}.

    \item {\bf Experimental result reproducibility}
    \item[] Question: Does the paper fully disclose all the information needed to reproduce the main experimental results of the paper to the extent that it affects the main claims and/or conclusions of the paper (regardless of whether the code and data are provided or not)?
    \item[] Answer: \answerYes{} 
    \item[] Justification: The implementation details are provided in Appendix~\ref{app:overall_algorithm}.

\item {\bf Open access to data and code}
    \item[] Question: Does the paper provide open access to the data and code, with sufficient instructions to faithfully reproduce the main experimental results, as described in supplemental material?
    \item[] Answer: \answerNo{} 
    \item[] Justification: We provide implementation details in Appendix~\ref{app:overall_algorithm} and will release the code upon acceptance of this paper.  

\item {\bf Experimental setting/details}
    \item[] Question: Does the paper specify all the training and test details (e.g., data splits, hyperparameters, how they were chosen, type of optimizer, etc.) necessary to understand the results?
    \item[] Answer: \answerYes{} 
    \item[] Justification: All the data processing, training and evaluation details are described in Section~\ref{sec:exp} and Appendix~\ref{sec:appendix}.

\item {\bf Experiment statistical significance}
    \item[] Question: Does the paper report error bars suitably and correctly defined or other appropriate information about the statistical significance of the experiments?
    \item[] Answer: \answerYes{} 
    \item[] Justification: We report the mean CTR with 95\% confidence interval in Figure~\ref{fig:vt_result}.

\item {\bf Experiments compute resources}
    \item[] Question: For each experiment, does the paper provide sufficient information on the computer resources (type of compute workers, memory, time of execution) needed to reproduce the experiments?
    \item[] Answer: \answerYes{} 
    \item[] Justification: The computational resources and training times are provided in Appendix~\ref{app:overall_algorithm}.
    
\item {\bf Code of ethics}
    \item[] Question: Does the research conducted in the paper conform, in every respect, with the NeurIPS Code of Ethics \url{https://neurips.cc/public/EthicsGuidelines}?
    \item[] Answer: \answerYes{} 
    \item[] Justification: We respect the NeurIPS Code of Ethics.

\item {\bf Broader impacts}
    \item[] Question: Does the paper discuss both potential positive societal impacts and negative societal impacts of the work performed?
    \item[] Answer: \answerNA{} 
    \item[] Justification: This work focuses on the technical foundations of sequential recommendation. 
    
\item {\bf Safeguards}
    \item[] Question: Does the paper describe safeguards that have been put in place for responsible release of data or models that have a high risk for misuse (e.g., pretrained language models, image generators, or scraped datasets)?
    \item[] Answer: \answerNA{} 
    \item[] Justification:  Our paper does not release data or models with a high risk for misuse.

\item {\bf Licenses for existing assets}
    \item[] Question: Are the creators or original owners of assets (e.g., code, data, models), used in the paper, properly credited and are the license and terms of use explicitly mentioned and properly respected?
    \item[] Answer: \answerYes{} 
    \item[] Justification: All the baselines are properly cited and introduced in Section~\ref{sec:exp}.

\item {\bf New assets}
    \item[] Question: Are new assets introduced in the paper well documented and is the documentation provided alongside the assets?
    \item[] Answer: \answerNA{} 
    \item[] Justification: The paper does not release new assets.

\item {\bf Crowdsourcing and research with human subjects}
    \item[] Question: For crowdsourcing experiments and research with human subjects, does the paper include the full text of instructions given to participants and screenshots, if applicable, as well as details about compensation (if any)? 
    \item[] Answer: \answerNA{} 
    \item[] Justification: The paper does not involve crowdsourcing or research with human subjects.

\item {\bf Institutional review board (IRB) approvals or equivalent for research with human subjects}
    \item[] Question: Does the paper describe potential risks incurred by study participants, whether such risks were disclosed to the subjects, and whether Institutional Review Board (IRB) approvals (or an equivalent approval/review based on the requirements of your country or institution) were obtained?
    \item[] Answer: \answerNA{} 
    \item[] Justification: Our paper does not involve crowdsourcing or research with human subjects.

\item {\bf Declaration of LLM usage}
    \item[] Question: Does the paper describe the usage of LLMs if it is an important, original, or non-standard component of the core methods in this research? Note that if the LLM is used only for writing, editing, or formatting purposes and does not impact the core methodology, scientific rigorousness, or originality of the research, declaration is not required.
    \item[] Answer: \answerNA{} 
    \item[] Justification: The core method development in this research does not involve LLMs. 

\end{enumerate}


\appendix

\section{Technical Appendices and Supplementary Material}
\label{sec:appendix}
\subsection{Derivations of Problem~\ref{eq:quantile_reg}}
\label{ref:derivation_of_P6}

In this section, we show that Problem~\ref{eq:IRL-initial} can be simplified and translated to Problem~\ref{eq:quantile_reg}, under the assumption that $Q(s,a)$ follows a quantile distribution $Z_\lambda(s,a)$. Given Problem~\ref{eq:IRL-initial}:

\begin{equation}
\max_{Q\in\Omega} \mathbb{E}_{s,a\sim\rho_E}[\psi(Q(s,a)-\gamma \mathbb{E}_{s'\sim P(s,a)}V^*(s'))]
 - (1-\gamma) \mathbb{E}_{s_0\sim \rho_0}[V^*(s_0)], \tag{4}
\end{equation}

The second term can be expanded as:

\begin{align}
\label{eq:v0}
   (1-\gamma) \mathbb{E}_{s_0\sim \rho_0}[V^*(s_0)]&=(1-\gamma)\sum_{t=0}^\infty\mathbb{E}_{D_E}[V^*(s_t)]-(1-\gamma)\sum_{t=1}^\infty\mathbb{E}_{D_E}[V^*(s_t)] \nonumber\\
   &=\mathbb{E}_{\rho_E}[V^*(s)-\gamma V^*(s')],
\end{align}

Since we already have the expert trajectory dataset $D_E$, we will use $\rho_E$ to estimate Equation~\ref{eq:v0}. Then, on substituting $\psi(x)=x-\frac{1}{4\alpha}x^2$ and $V^*(s)=\log\sum_a\exp Q(s,a)$ in Problem~\ref{eq:IRL-initial}, we have:
\begin{equation}
    \max_{Q} \mathbb{E}_{\rho_E}[Q(s,a)-\log \sum_a \exp Q(s,a)]
- \frac{1}{4\alpha}\mathbb{E}_{\rho_E}[(Q(s,a)-\log \sum_a \exp Q(s',a))^2].
\end{equation}
As $Q(s,a)$ is parameterized by quantiles $\{\lambda_i(s,a)\}_{i=1}^N$, we replace $Q$ with $Q_\lambda$ and obtain Problem~\ref{eq:quantile_reg}.

\subsection{Optimization steps for QR-IQL}
\label{ref:qr-iql}
In order to learn the distributional $Q$-function, we change the output layer of the $Q$-network to be of size $|A|\times N$, where $|A|$ denotes the size of the action space and $N$ denotes the number of quantiles. For a given action, each of the $N$ heads implicitly correlates to a $\lambda_i$. We adopt the widely used Pinball loss to learn the positions of $\{\lambda_i\}_{i=1}^N$, which is defined as:
$$ p_\lambda(u)=\left\{
\begin{aligned}
&\lambda\cdot u,  & \quad u \geq 0 \\
&(\lambda-1)\cdot u, & \quad u<0
\end{aligned}
\right.
$$
where $u$ represents the error between the predicted value and the target value at quantile $\lambda$. Intuitively, the Pinball loss pushes the quantile $\lambda$ to the right position so that the predicted value distribution matches the target distribution. Based on the objective of Problem~\ref{eq:quantile_reg}, we define two errors as: 
$$u^1_{\lambda_i}(s,a)=\lambda_i(s,a)-\log \sum_a \exp \lambda_i(s,a),$$ 
$$u^2_{\lambda_i}(s,a,s')=\frac{1}{4\alpha}(\lambda_i(s,a)-\log \sum_a \exp \lambda_i(s',a))^2,$$
where $\lambda_i$ is the $i$-th quantile and $\lambda_i(s,a)$ is the corresponding head of the $Q_\lambda$-network. Complete training procedures are provided in Algorithm~\ref{alg:qr_iql}.

\begin{algorithm}
\caption{QR-IQL Optimization Steps}
\label{alg:qr_iql}
\textbf{Input}: Interaction (expert) data $D_E$, number of quantiles $N$

\begin{algorithmic}[1] 
\STATE Initialize network $Q_\lambda$;\\
\REPEAT
\STATE Sample (batched) data $(s,a,s')$;\\
\STATE Compute errors $u^1_{\lambda_i}(s,a)$ and $u^2_{\lambda_i}(s,a,s')$ for each quantile in $\{\lambda_i\}_{i=1}^N$;\\
\STATE Compute the Pinball losses $p_{\lambda_i}(u^1)$ and $p_{\lambda_i}(u^2)$ for each quantile in $\{\lambda_i\}_{i=1}^N$;\\
\STATE Compute the total loss as: $\sum_{i=1}^N[p_{\lambda_i}(u^1) + p_{\lambda_i}(u^2)]$;\\
\STATE Minimize the total loss by Adam \cite{adam};
\UNTIL{convergence}
\end{algorithmic}
\textbf{Output}: $Q^*_{\lambda}$
\end{algorithm}

\subsection{Implementation Details}
\label{app:overall_algorithm}
Our experiments are run on a server with 2×AMD EPYC 7542 32-Core Processor CPU and 2×NVIDIA RTX 3090 graphics. For the offline experiments on Amazon datasets, it takes about 3 hours for  50,000 iterations of training with a 4000 batch size. For online experiments on Virtual Taobao, it takes about 4 hours for 50,000 RL training steps. 

Algorithm~\ref{alg:algorithm} describes an overview of the implementation procedures. Basically, there are two stages. At stage 1 we focus on learning $Q_{\lambda}^*$ and at stage 2 we focus on learning $F_{\zeta}^*$. In practice, the architecture of the recommendation model $F_{\zeta}$ could be of various types. For example, in Section \ref{sec:offline_exp}, we test seven widely used recommendation models: Wide\&Deep \cite{widedeep}, PNN \cite{pnn}, DeepFM \cite{deepfm}, DIN \cite{din}, DIEN \cite{dien}, LinRec \cite{linrec} and SIGMA \cite{sigma}. We will use the same network architecture of $F_{\zeta}$ to construct $Q_{\lambda}$ (except the output layer) to ensure that the features are processed properly. All the hyper-parameters of the backbone models follow their official codes. For implementation of MTRec, we select the number of quantiles $N=10$ and the weight $\alpha=0.5$ in Problem~\ref{eq:quantile_reg}. 

Actually, our MTRec framework is industrial friendly due to the following merits. First, all the training procedures are run in an offline manner, saving the cost of building online user-system interaction environments. Second, since $Q_{\lambda}$ shares the model architecture with $F_{\zeta}$, we do not need to build a mental reward model from scratch, saving a lot of work for adaptation.

\begin{algorithm}
\caption{Overall Implementation of MTRec}
\label{alg:algorithm}
\textbf{Input}: Interaction (expert) data $D_E$

\begin{algorithmic}[1] 
\STATE Initialize networks $Q_\lambda$ and $F_\zeta$;
\STATE Learn $Q_\lambda^*$ according to Algorithm~\ref{alg:qr_iql};
\STATE Obtain the mental rewards by \\
$r(s,a)\leftarrow Q_\lambda^*(s,a)-\gamma V^*(s')$, $\forall s,a\sim D_E$;
\STATE Patch the mental rewards to $D_E$;
\STATE Learn $F_\zeta^*$ according to Equation~\ref{eq:f1} or Equation~\ref{eq:f2};
\end{algorithmic}
\textbf{Output}: Aligned Recommendation Model $F^*_{\zeta}$
\end{algorithm}
\subsection{Details on the Amazon datasets}
\label{seq:stats_amazon}
The Amazon dataset \cite{amazon} collects user review data from amazon.com. The crawled reviews have a time span from May 1996 to July 2014. The dataset can be divided into many subsets according to the various product categories. To verify the effectiveness of MTRec, we utilized two subsets of the Amazon dataset: Books and Electronics. We treated the reviews as user behaviors and sorted the reviews from each user chronologically. Based on a user's historical behaviors, our goal was to predict whether the user would write a review.

\begin{table}[htbp]
    \centering
    \begin{tabular}{c c c c c}
    \toprule
     Dataset & User & Item & Categoriy & Sample \\
    \midrule
    Books & 603,668 & 367,982 & 1,600 & 603,668 \\
    Electronics & 192,403 & 63,001 & 801 & 192,403 \\
    \bottomrule\\
    \end{tabular}
    \caption{The statistics of the Amazon datasets.}
    \label{table:dataset}
\end{table}

\subsection{Limitations}
\label{sec:limitation}
While our work opens a door for studying users' intrinsic rewards during their interaction with the recommender systems, it still lacks a systematic method to thoroughly evaluate the learned mental reward model. In future works, we plan to construct a comprehensive benchmark involving large-scale human studies to further evaluate the mental reward model. 


\end{document}